# Program Optimization Based Pointer Analysis and Live Stack-Heap Analysis

**Mohamed A. El-Zawawy**

**Department of Mathematics, Faculty of Science, Cairo University**
**Giza, 12316, Egypt**

### Abstract

In this paper, we present type systems for flow-sensitive pointer analysis, live stack-heap (variables) analysis, and program optimization. The type system for live stack-heap analysis is an enrichment of that for pointer analysis; the enrichment has the form of a second component being added to types of the latter system. Results of pointer analysis are proved useful via their use in the type system for live stack-heap analysis. The type system for program optimization is also an augmentation of that for live stack-heap analysis, but the augmentation takes the form of a transformation component being added to inference rules of the latter system. The form of program optimization being achieved is that of dead-code elimination. A form of program correction may result indirectly from eliminating faulty code (causing the program to abort) that is dead. Therefore program optimization can result in program correction. Our type systems have the advantage of being compositional and relatively-simply structured.

The novelty of our work comes from the fact that our type system for program optimization associates the optimized version of a program with a justification (in the form of a type derivation) for the optimization. This justification is pretty much appreciated in many research areas like certified code (proof-carrying code) which is the motivation of this work.

***Keywords***: *Pointer analysis, Live variables analysis, Live stack-heap analysis, Program optimization, Type systems, Certified code.*

## 1. Introduction

Rather than dynamic code analysis concerned with analyzing programs during execution time, static code analysis (statics analysis) [14] is a concept describing analyzing programs without actually executing them. Static analysis can result in improving the quality of the code in different ways (including correcting and optimizing the code) or in verifying industrial standards of the code. Data-flow analysis, one of the techniques used in static analysis, is useful for collecting information for each program point. An analysis whose results do not change due to permuting a statement sequence $S_1;S_2$ into $S_2;S_1$ is described as flow-insensitive; otherwise it is described as flow-sensitive. For a flow-sensitive analysis, if the program is traversed forwardly (backwardly) to collect information, the technique is called a forward (backward) analysis. If the collected information may (must) be true, the technique is described as may (must). Examples of forward-may and backward-may analyses are pointer and live variables analyses [14], respectively. Pointer analysis calculates for each store (a variable or a memory location) at every program point the set of addresses that have a chance of being contained in that store at that program point. Roughly speaking, live variables analysis calculates for every program point the set of variables used later in the program. In case of pointer programs, we call this analysis live stack-heap analysis and it calculates the set of variables and memory locations that are used later in the program. Results of live variables analysis can be used to eliminate unnecessary code in a technique called dead-code elimination.

Although static analysis is usually treated in an algorithmic style [14], there are other frameworks that can be used to successfully achieve static analysis. One of such frameworks is type systems [9, 20, 2, 15] that has proved so far to be a very convenient tool for this job. In the algorithmic fashion, the work is done on an annotated form of the program control- flow graph. However in the type-systems manner the work is done on the program using its phrase structure. This fact is advantageous to the use of the type-systems framework when it comes to optimizing programs. This is so because the algorithmic style usually produces only the optimized version of the program. However the type-systems style is conveniently-capable of producing the optimized version together with a justification (in the form of a type derivation) for the optimization. This justification is necessary in applications like certified code. Also the relative simplicity of inference rules of type systems makes their framework auspicious.

**Motivation**

The program on the left-hand side of Figure 1 is a motivating example of our work. Suppose that $y$ is the only variable whose value concerns us at the end of the program. Then the last statement is unnecessary (dead code). Also the assignment statement in line 6 is a dead code and it causes the program to aborts because the value of $i$ is not in the domain of the heap. Therefore removing theses statements optimizes the program and in the same time removes a cause for abortion.

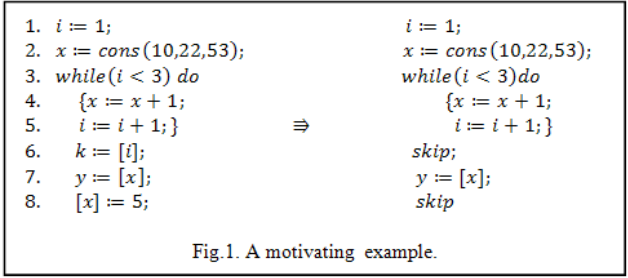

Fig.1. A motivating example.

The objective of this paper is to develop a technique that transforms a program like this one into an optimized version like that in the right-hand side of Figure 1 and also produces a proof or justification for the transformation process.

All in all this paper tackles the problem of transforming pointer programs into optimized and possibly corrected versions and producing justifications for the transformation process. The importance of producing the justification comes from the area of certified code which is the motivation of our work and which provides good applications for the work as well. The program optimization, meant here, is dead-code elimination. The optimized version of a program is possibly a corrected version as well; this is the case if reasons for abortions in the program are included in dead code and hence gets removed with the dead code. In other words, program optimization can result in program correction. Our tool for solving this problem is type systems. Up to our knowledge, our paper is the first to tackle this problem (using type systems) for pointer programs.

**Contributions**

1. An original type system for flow-sensitive pointer analysis.
2. A novel type system for live stack-heap (variables) analysis for pointer programs. This type system utilizes results of our type system for pointer analysis and is an enrichment of it.
3. The third contribution is a new type system for optimizing and possibly correcting pointer programs. This type system serves also as a tool for obtaining a justification (in the form of a type derivation) for every individual transformation and is an augmentation of our type system for live stack-heap analysis.

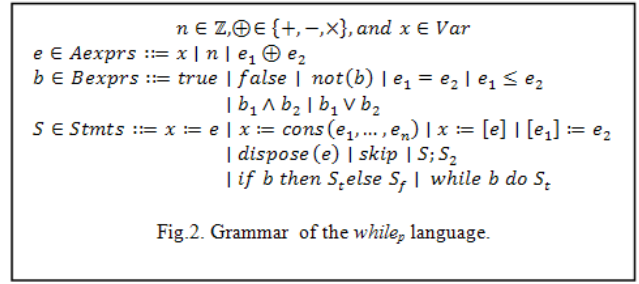

Fig.2. Grammar of the $while_p$ language.

**Organization**

The rest of the paper is organized as follows. The language $while_p$ (the while language enriched with pointer commands) and an operational semantics for its constructs are presented in Section 2. Our proposed type systems for flow-sensitive pointer analysis and live stack-heap analysis are presented in Sections 3 and 4, respectively. The type system carrying program optimization is introduced in Section 5. A brief survey of related work is presented in Section 6.

## 2. The Programming Language

The programming language that we are using is usually used to introduce separation logic like in [19] and its operational semantics is a slightly-modified version of that in [19]. The language is an imperative one that is enriched with commands dealing with pointers. We call this language $while_p$. This section presents the language $while_p$ with an operational semantics to its constructs. The grammar of the $while_p$ language is shown in Figure 2, where $Var$ is a finite set of program variables.

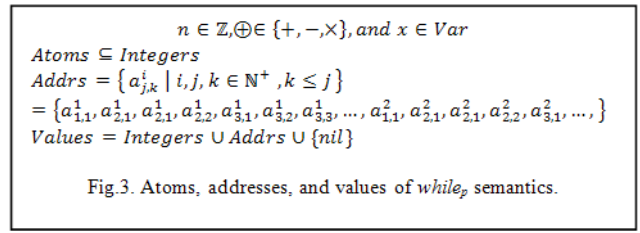

Fig.3. Atoms, addresses, and values of $while_p$ semantics.

For any $m \in \mathbb{N}^+$, we assume that the memory has an infinite number of arrays of length $m$ with addresses $\{a^1_{m,1}, a^1_{m,2}, \ldots, a^1_{m,l}, a^2_{m,1}, a^2_{m,2}, \ldots, a^2_{m,m}, \ldots\}$ Therefore the set of address, $Addrs$, has the form presented in Figure 3. This memory model, rather than letting addresses to be a subset of integers, has the advantage of reducing the chance of messing with the memory. This is so because a number which is intended to be used as a numerical value (not as an address) can be an address as well and therefore it can be accidently used to access un-allowed or unintended memory cells. In order to facilitate evaluating inequalities we assume that $Values$ is equipped with an order.

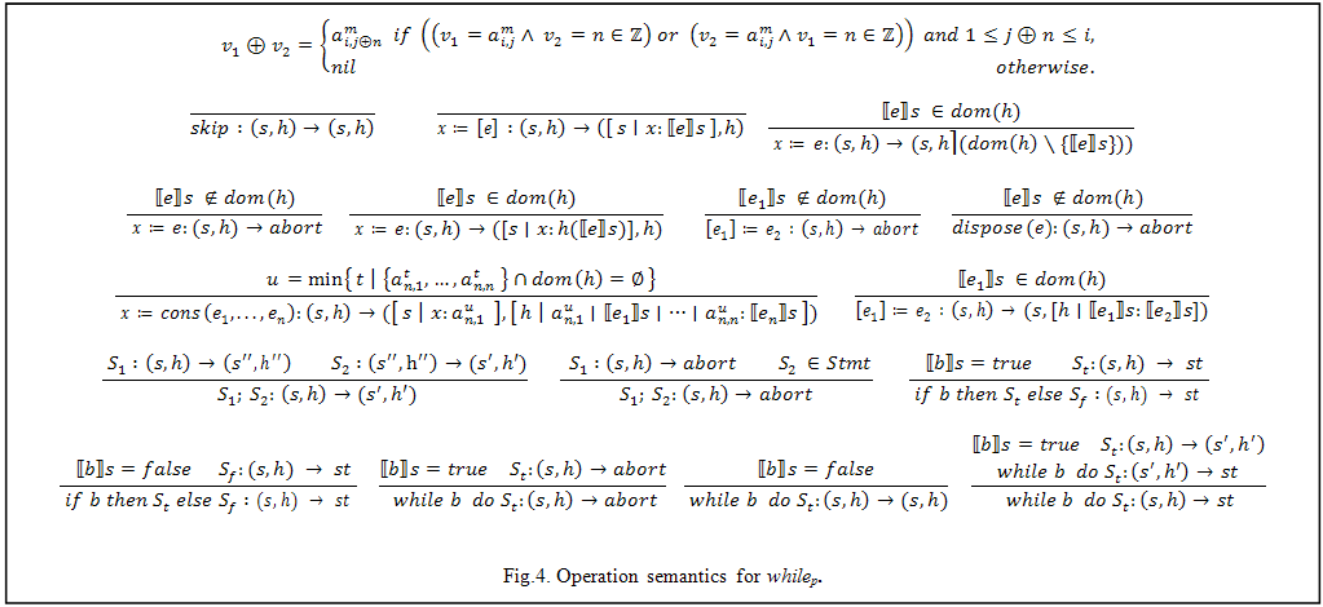

Fig.4. Operation semantics for $while_p$.

**Definition 1.**
A stack (heap) is a map (finite partial map) from Var (Addrs) to Values. A state is an abort or a pair of a stack and a heap.

Arithmetic and Boolean expressions have the same semantics as in the case of the *while* language except for the operation $\oplus$. The semantics of this operation on *Values* is defined as usual if both of its operands are integers and otherwise as in Figure 4.

The semantics of *whilep* statements is given by an operational semantics whose transition relation is denoted by $\rightarrow$ and whose configurations (nonterminal and terminal) are defined in Definition 1. In the inference rules of the semantics (Figure 4), *st* denotes a state.

The *cons* allocates the array $a^u_{n,1}, \ldots, a^u_{n,n}$ with the minimum dimension, $u$, of all available arrays of length $n$. The allocation takes place by storing the address $a^u_{n,1}$ in $x$ and semantics of expressions $e_1, \ldots, e_n$ in addresses $a^u_{n,1}, \ldots, a^u_{n,n}$, respectively. If $f$ is a map and $A$ is a set, $f \rceil A$ denotes the restriction of $f$ on $A$ and $[f \mid x{:}A]$ denotes the function whose domain is $dom(f) \cup \{x\}$ and whose definition is $\lambda y.$ if $y = x$ then $A$ else $f(y)$.

## 3. Pointer Analysis

In this section, we introduce a type system for flow-sensitive pointer analysis which is a forward-may analysis that assigns to each program point a partial map from variables and memory locations to the power set of addresses. Under this map, the image of an element is an over-approximate set of addresses that the element may contain (point to) at this program point. The set of points-to types, *PTS*, and the sub-typing relation are defined as follows.

**Definition 2.**
1. $PTS = \{pts \mid pts{:} Var \cup d \rightarrow 2^{Addrs} \mid d \subseteq Addrs\}$. The bottom type is denoted by $\bot$.
2. $pts \leq pts' \Leftrightarrow dom(pts) \subseteq dom(pts')$ and $\forall t \in dom(pts). pts(t) \subseteq pts'(t)$.
3. A state $(s, h)$ has type *pts*, denoted by $(s, h) \models pts$, if
   – $dom(h) \subseteq dom(pts)$,
   – $\forall x \in Var. s(x) \in Addrs \Rightarrow s(x) \in pts(x)$, and
   – $\forall a \in dom(h). h(a) \in Addrs \Rightarrow h(a) \in pts(a)$.

Given a points-to type *pts*, the pointer analysis is achieved for a statement $S$ via a post-type derivation for $S$ and *pts* as the pre-type. Typically the pre-type *pts* is the bottom type $\bot$. The judgment of an arithmetic expression $e$ has the form $e{:} pts \rightarrow V$. The set $V$ is either a set of addresses or a singleton of an integer. The intended meaning, which is formalized in Lemma 1, of this judgment is that $V$ captures any address that $e$ evaluates to in a state of type *pts*. In particular if $V$ is a set of addresses, then $e$ is either an address from $V$, any integer, or *nil*. The judgment of a statement $S$ has the form $S{:} pts \rightarrow pts'$. The intuition, which is formalized in Theorem 1, of this judgment is that if $S$ is executed in a state of type *pts*, then any state (rather than *abort*) where the execution ends is of type *pts'*. In the rest of the paper when $e{:} pts \rightarrow V$, we let V' denotes $V \cap Addrs$. The inference rules of our type system for pointer analysis are presented in Figure 5.

**Lemma 1.**
Suppose that $(s, h) \models pts$ and $e{:} pts \rightarrow V$. Then

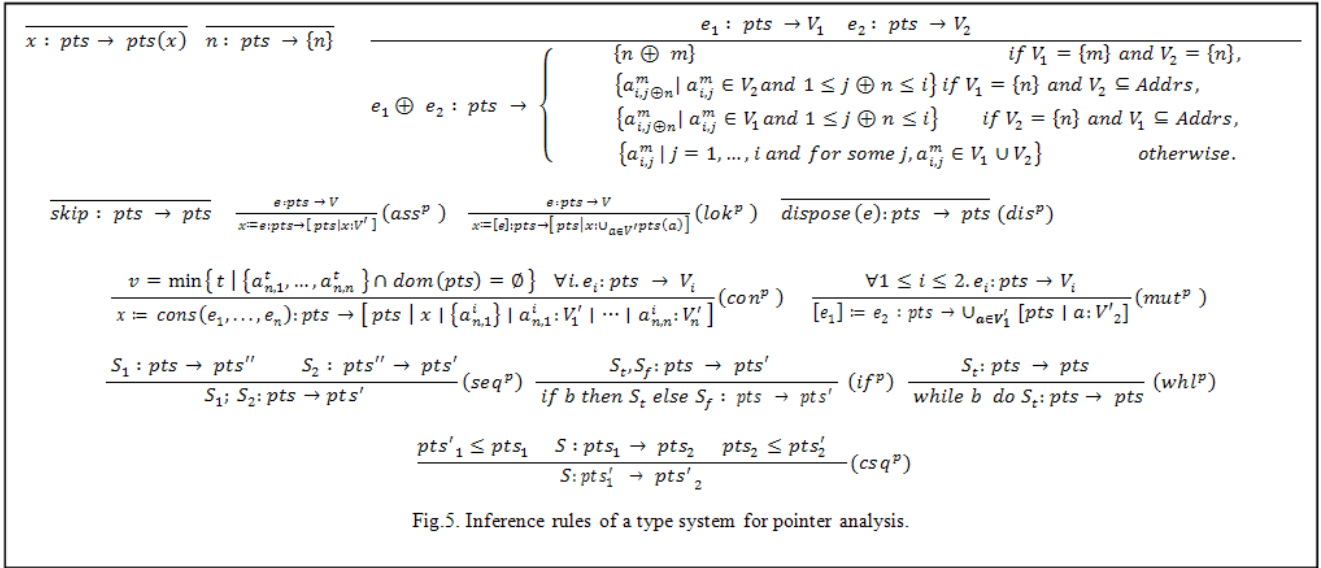

Fig.5. Inference rules of a type system for pointer analysis.

1. $V \subseteq Addrs$ or $\exists n \in Z. \ V = \{n\}$,
2. $\forall n \in Z. \ V = \{n\} \Rightarrow [\![e]\!]s = n$, and
3. $[\![e]\!]s \in Addrs \Rightarrow [\![e]\!]s \in V$.

**Proof.** By induction on the structure of $e$. We present the proof of the last item. If $e = n$ then $[\![e]\!]s = n \notin Addrs$. If $e = x$ then $[\![x]\!]s = s(x) \in Addrs$ implies $s(x) \in pts(x) = V$ because $(s, h) \models pts$. Now suppose $e = e_1 \oplus e_2$, $e_1 : pts \to V_1$, and $e_2 : pts \to V_2$. If $[\![e_1 \oplus e_2]\!]s \in Addrs$, then we have one of the following cases:
1. $[\![e_1]\!]s = a^m_{i,j}$, $[\![e_2]\!]s = n$, and $1 \leq j \oplus n \leq i$.
2. $[\![e_1]\!]s = n$, $[\![e_2]\!]s = a^m_{i,j}$, and $1 \leq j \oplus n \leq i$.

In the first case, by the induction hypothesis $a^m_{i,j} \in V_1$ and if $V_2 = \{t\}$, then, by (2), $n = t$ and $[\![e_1 \oplus e_2]\!]s = a^m_{i,j \oplus n} \in \{a^m_{i,j \oplus n} \mid a^m_{i,j} \in V_1 \land 1 \leq j \oplus n \leq i\} = V$. If $V_2 \subseteq Addrs$ then $[\![e_1 \oplus e_2]\!]s = a^m_{i,j \oplus n} \in V$. The second case is similar to the first case. ∎

The following lemma is needed in the proof of the following soundness theorem and it is obvious because $(s, h) \models pts$ implies $dom(h) \subseteq dom(pts)$.

**Lemma 2.**
*Suppose that $(s, h) \models pts$, $u = min\{t \mid \{a^t_{n,1}, \ldots, a^t_{n,n}\} \cap dom(h) = \emptyset\}$, and $v = min\{t \mid \{a^t_{n,1}, \ldots, a^t_{n,n}\} \cap dom(pts) = \emptyset\}$. Then $1 \leq u \leq v$.*

The rules ($ass_p$) and ($dis_p$) are straightforward. For the rule ($con_p$) and by Lemma 2, executing the *cons* statement in a state of type *pts* results in allocating one of the arrays $\{a^j_{n,1}, \ldots, a^j_{n,n}\}$, $1 \leq j \leq v$. But it is not obvious which of these arrays will be allocated. Therefore the rule ($con_p$) takes into account all these possibilities by adding the addresses of these arrays to $pts(x)$ and adding $V'_i$ to the image, under *pts*, of each location $a^j_{n,i}$.

In the rule ($lok_p$), $V'$ contains any address that $e$ evaluates to in a state of type *pts*. Therefore the set $\cup_{a \in V'} pts(a)$ captures any address that goes into $x$ after executing the look-up statement in a state of type *pts*. For the rule ($mut_p$), there are two cases for $V_1$, namely either $|V_1| = 1$ or $|V_1| \neq 1$. In the first case, the rule ($mut_p$) cuts down to a form that is pretty much similar to the rule ($ass_p$). In the second case, it is not obvious to which address the assignment will happen. Hence the post-type is calculated from the pre-type by including the set $V'_2$ in the image of every element of $V'_1$. The rules ($seq_p$), ($if_p$), and ($csq_p$) are clear.

As it is evident from the ($whl_p$) rule, an invariant type is necessary to type a while statement. The required invariant type is calculated as a fix-point of an order-preserving map over the complete lattice *pts* using a given pre-type. The consequence rule can be used to show that the fix-point is indeed the required invariant type.

**Theorem 1.** (*Soundness*)
1. $pts \leq pts'$ iff (For every state $(s, h)$, $(s, h) \models pts \Rightarrow (s, h) \models pts'$).
2. Suppose that $S: pts \to pts'$ and $S: (s, h) \to (s', h')$. Then $(s, h) \models pts$ implies $(s', h') \models pts'$.

**Proof.** 1. The left-to-right direction is obvious. The other direction is proved as follows. Suppose $x \in Var$, $a, b \in Addrs$, and $a \in pts(x)$. Then the state $(s, h) = (\{(x, a), (y, 0) \mid x \neq y \in Var\}, \emptyset)$ is of type *pts* and therefore is of type *pts'*. So $a \in pts'(x)$ and hence $pts(x) \subseteq pts'(x)$. Similarly, we can show that $b \in dom(pts)$ implies $b \in dom(pts')$ and $pts(b) \subseteq pts'(b)$.

2. The proof is by induction on the structure of type derivation as follows:
(a) The type derivation has the form ($ass_p$). In this case, $pts' = [pts \mid x:V']$ and $(s', h') = ([s \mid x : [\![e]\!]s], h)$. If $[\![e]\!]s \in Addrs$, then $[\![e]\!]s \in V'$ by Lemma 1. Therefore $s'(x) \in Addrs$ implies $s'(x) \in pts'(x)$. We also have that $dom(h') = dom(h) \subseteq dom(pts) \subseteq dom(pts')$ because $(s, h) \models pts$. It is obvious that for any $x \neq y \in Var$ and $a \in dom(h')$, $s'(y) \in Addrs$ implies $s'(y) \in pts'(y)$ and $h'(a) \in Addrs$ implies $h'(a) \in pts'(a)$. Hence $(s', h') \models pts'$.

(b) The type derivation has the form ($lok_p$). In this case, $pts' = [pts \mid x : \cup_{a \in V'} pts(a)]$ and $(s', h') = ([s \mid x : h(\llbracket e \rrbracket s)], h)$. Also we have $\llbracket e \rrbracket s \in Addrs \cap dom(h)$ and hence $\llbracket e \rrbracket s \in V'$ by Lemma 1. If $h(\llbracket e \rrbracket s) \in Addrs$, then $h(\llbracket e \rrbracket s) \in pts(\llbracket e \rrbracket s)$ because $(s, h) \models pts$. Therefore $s'(x) = h(\llbracket e \rrbracket s) \in Addrs$ implies $s'(x) = h(\llbracket e \rrbracket s) \in \cup_{a \in V'} pts(a) = pts'(x)$. We also have that $dom(h') = dom(h) \subseteq dom(pts) \subseteq dom(pts')$ because $(s, h) \models pts$. It is obvious that for any $x \neq y \in Var$ and $a \in dom(h')$, $s'(y) \in Addrs$ implies $s'(y) \in pts'(y)$ and $h;(a) \in Addrs$ implies $h'(a) \in pts'(a)$. Hence $(s', h') \models pts'$

(c) The type derivation has the form ($con_p$). In this case, $pts' = \cup_{1 \leq i \leq v} [pts \mid x : \{a^i_{n,1}\} \mid a^i_{n,1} : V'_1 \mid \ldots \mid a^i_{n,n} : V'_n]$ and $(s', h') = ([s \mid x : a^u_{n,1}], [h \mid a^u_{n,1} : \llbracket e_1 \rrbracket s \mid \ldots \mid a^u_{n,n} : \llbracket e_n \rrbracket s])$. By Lemma 2, $1 \leq u \leq v$. For every $1 \leq i \leq n$, if $\llbracket e_i \rrbracket s \in Addrs$ then $\llbracket e_i \rrbracket s \in V'_i$ by Lemma 1. We have $s'(x) = a^u_{1,n} \in \{a^1_{1,n}, \ldots, a^v_{1,n}\} \subseteq pts'(x)$. We also have that $dom(h') \subseteq dom(pts')$ because $dom(h) \subseteq dom(pts)$ ($(s, h) \models pts$) and $1 \leq u \leq v$. It is obvious that for any $x \neq y \in Var$ and $a \in dom(h') \setminus \{a^u_{n,1}, \ldots, a^u_{n,n}\}$, $s'(y) \in Addrs$ implies $s'(y) \in pts'(y)$ and $h'(a) \in Addrs$ implies $h'(a) = h(a) \in pts(a) \subseteq pts'(a)$. For every $1 \leq i \leq n$, if $h(a^u_{n,i}) \in Addrs$, then $h(a^u_{n,i}) = \llbracket e_i \rrbracket s \in V'_i \subseteq pts'(a^u_{n,i})$. Hence $(s', h') \models pts'$.

(d) The type derivation has the form ($mut_p$). In this case, $pts' = \cup_{a \in V'_1} [pts \mid a : V'_2]$ and $(s', h') = (s, [h \mid \llbracket e_1 \rrbracket s : \llbracket e_2 \rrbracket s])$. We have $\llbracket e_1 \rrbracket s \in dom(h) \cap V_1$ and if $\llbracket e_2 \rrbracket s \in Addrs$ then $\llbracket e_2 \rrbracket s \in V'_2$ by Lemma 1. If $h'(\llbracket e_1 \rrbracket s) \in Addrs$, then $h'(\llbracket e_1 \rrbracket s) = \llbracket e_2 \rrbracket s \in V'_2 \subseteq pts'(\llbracket e_1 \rrbracket s)$ because $\llbracket e_1 \rrbracket s \in V'_1$. We also have that $dom(h') = dom(h) \subseteq dom(pts) \subseteq dom(pts')$ because $(s, h) \models pts$. It is obvious that for any $y \in Var$ and $a \in dom(h') \setminus V'_1$, $s'(y) \in Addrs$ implies $s'(y) = s(y) \in pts(y) = pts'(y)$ and $h'(a) \quad Addrs$ implies $h'(a) = h(a) \in pts(a) \quad pts'(a)$. Hence $(s', h') \models pts'$.

The remaining cases are straightforward to check. ∎

## 4. Live stack-heap analysis

In this section, we show how the type system for pointer analysis, presented in the previous section, can be enriched to produce a type system for live stack-heap analysis. In other words, the type system presented in this section is a strict extension of the system presented in the previous section. This reflects the fact that pointer information obtained by previous system are used to improve the precision of the live stack-heap analysis.

The live stack-heap analysis associates with each program point the set of variables and memory locations live (according to Definition 3) at that point. The resulting type system is a generalized one of that presented in [20] for live variables analysis for the while language. Therefore the goal in this section is to utilize results of our type system for pointer analysis and to build o n it a type system that carries live stack-heap analysis. Towards this objective, we augment points-to types to get live stack-heap types defined below (Definition 4).

**Definition 3.**
- *A* variable (memory location) *is live at a program point if there is a computational path from that program point during which the variable (the memory location's content) gets usefully used before being modified.*
- *A variable (the content of a memory location) is* usefully used
    1- *if it is used in an assignment to a variable or a memory location that is live at the end of the assignment,*
    2- *the guard of an if-statement or a while-statement,*
    3- *an arithmetic expression of a dispose statement, or*
    4- *the left expression of a mutation statement.*

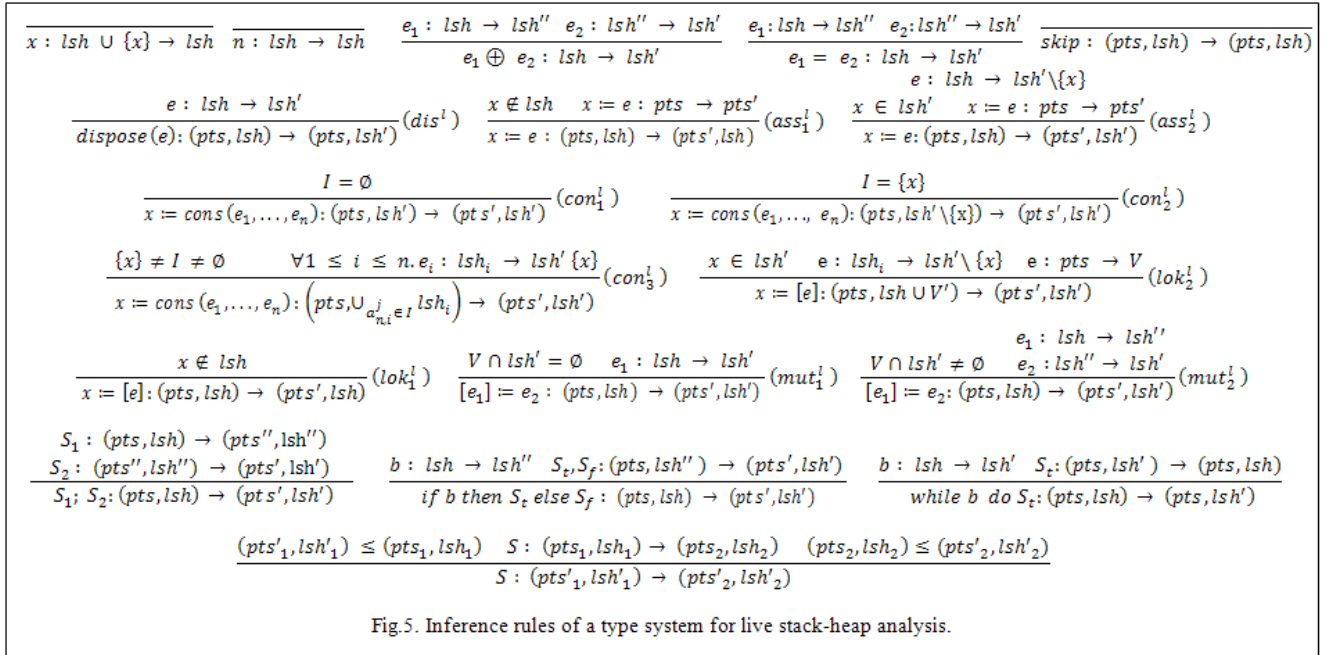

Fig.5. Inference rules of a type system for live stack-heap analysis.

**Definition 4.**

*The set of live stack-heap types (live types in short) is denoted by lsh and equal to $pts \times 2^{Var \cup Addrs}$. The second component of a live type is termed a live-component. The subtyping relation $\leq$ is defined as:*

$(pts, lsh) \leq (pts', lsh')$ iff $(pts \leq pts'$ and $lsh \supseteq lsh')$.

The judgment of an expression has the form $e : lsh \to lsh'$ and the intuition is that $lsh$ is $lsh'$ plus variables occurring free in $e$. The judgment of a statement $S$ has the form $S : (pts, lsh) \to (pts', lsh')$ and it is meant to guarantee that if $lsh'$ contains variables and memory locations live at the post-state of an execution of $S$, then $lsh$ contains variables and memory locations live at the pre-state of this execution. This is formalized in Theorem 2 and consents with the fact that live stack-heap analysis is a backward analysis. This also gives an insight into the definition of $(s, h) \models lsh$ in Definition 6.

Suppose that we have a live-component $lsh'$ and the result of a pointer analysis for a statement $S$ (in the form $S : pts \to pts'$). The live stack-heap analysis is achieved for $S$ via a pre-type derivation that calculates a set $lsh$ such that $S: (pts, lsh) \to (pts', lsh')$. It is natural to let $lsh'$ be the set of variables that we have interest in their values at the end of executing $S$.

The inference rules for our type system for live stack-heap analysis are presented in Figure 5. The inference rules for Boolean expressions other than $e_1 = e_2$ are similar to the inference rule for $e_1 = e_2$. In rules for allocation, we suppose that $x := cons(e_1, \ldots, e_n) : pts \to pts'$, $v = min\{t \mid \{a^t_{n,1}, \ldots, a^t_{n,n}\} \cap dom(pts) = \emptyset\}$, and $I = \{x, a^i_{n,1}, \ldots, a^i_{n,n} \mid 1 \leq i \leq v\} \cap lsh'$. In rules for look-up, we assume that $x := [e] : pts \to pts'$. For the mutation statement, we suppose that $[e_1] := e_2 : pts \to pts'$ and $e_1 : pts \to V$.

The set of variables and memory locations modified by the statement $x := cons(e_1, \ldots, e_n)$ is contained in $\{x, a^1_{n,1}, \ldots, a^1_{n,n}, a^2_{n,1}, \ldots, a^2_{n,n}, \ldots, a^v_{n,1}, \ldots, a^v_{n,n}\}$. We have three cases concerning which elements of this set are in $lsh'$ (possibly live after executing the statement); none, only $x$, or at least one address. For the first case when all modified elements are necessary dead after executing the statement, the rule $(cons^l_1)$ equalizes live-components of the pre and post types. For the second case, the rule $(cons^l_2)$ lets $lsh' \setminus \{x\}$ (as the assignment to $x$ kills it) be the live-component of the pre-type. For the third case treated by the rule $(cons^l_3)$ the live-component of the pre-type is constructed via augmenting $lsh' \setminus \{x\}$ with variables occurring free in every expression assigned to a location possibly live after execution.

For the look-up statement, the rule $(lok^l_2)$ deals with the case that $x$ is possibly live after executing the statement. In this case, the pointer information is used to calculate the set of addresses $V'$. Then, to form the live-component of the pre-type, the set $lsh' \setminus \{x\}$ is augmented to include $V'$ and variables occurring free in $e$.

For the mutation statement, the pointer information is used to find the set $V'$ containing any address that the expression $e_1$ evaluates to in a state of type $pts$. The type system has two rules dealing with the two possible cases; whether or not $V'$ has an empty intersection with $lsh'$. The rule $(mut^l_2)$ takes care of the case of nonempty intersection. In this rule the live-component of the pre-type is constructed by adding variables occurring free in $e_1$ and $e_2$ to $lsh'$. We note that in this case it is not obvious which location will be modified (and hence gets killed) but it is clear that this location is possibly live at the end of mutation. Therefore nothing is removed from $lsh'$; instead variables occurring free in $e_1$ and $e_2$ are added.

Now we introduce necessary definitions and results towards proving the soundness of our type system for live stack-heap analysis.

**Definition 5.** *1. $(s, h) \models_{lsh} pts \Leftrightarrow dom(h) \subseteq dom(pts)$, $\forall x \in Var \cap lsh$ $(s(x) \in Addrs \Rightarrow s(x) \in pts(x))$, and $\forall a \in dom(h) \cap lsh$ $(h(a) \in Addrs \Rightarrow h(a) \in pts(a))$.*
*2. $(s, h) \sim_{lsh} (s', h') \Leftrightarrow \forall x \in lsh \cap Var. s(x) = s'(x)$, and $\forall a \in lsh \cap dom(h) \cap dom(h'). h(a) = h'(a)$.*
*3. $(s, h) \sim_{(pts,lsh)} (s', h') \Leftrightarrow dom(h) = dom(h')$, $(s, h) \models_{lsh} pts$, $(s', h') \models_{lsh} pts$, and $(s, h) \sim_{lsh} (s', h')$.*

**Definition 6.** *The expression $(s, h) \models lsh$ denotes the case when there is a variable or a memory location that is live at that state (computational point) and is not included in lsh. A state $(s,h)$ has type $(pts,lsh)$, denoted by $(s,h) \models (pts, lsh)$, if $(s, h) \models_{lsh} pts$ and $(s, h) \models lsh$.*

The following lemma is proved by structure induction on *e* and *b*.

**Lemma 3.** *Suppose that $(s,h)$ and $(s',h')$ are states and lsh and $lsh' \in 2^{Var \cup Addrs}$. Then*
*1. If $lsh \supseteq lsh'$ and $(s,h) \sim_{lsh} (s',h')$, then $(s,h) \sim_{lsh'} (s',h')$.*
*2. If $e : lsh \to lsh'$ and $(s, h) \sim_{lsh} (s',h')$, then $[\![e]\!]s = [\![e]\!]s'$ and $(s, h) \sim_{lsh'} (s', h')$.*
*3. If $b : lsh \to lsh'$ and $(s,h) \sim_{lsh} (s',h')$, then $[\![b]\!]s = [\![b]\!]s'$ and $(s,h) \sim_{lsh'} (s', h')$.*

The following lemma follows from Lemma 1.

**Lemma 4.** *Suppose that $(s, h) \models_{lsh} pts$, $FV(e) \subseteq lsh$, and $e : pts \to V$. Then $[\![e]\!]s \in Addrs \Rightarrow [\![e]\!]s \in V$.*
*Proof.* Consider the state $(s',h')$, where $s' = \lambda x.$ if $x \in FV(e)$ then $s(x)$ else $0$ and $h' = \emptyset$. It is not hard to see that $[\![e]\!]s = [\![e]\!]s'$ and $(s',h') \models pts$. Now by Lemma 1, $[\![e]\!]s' \in Addrs$ implies $[\![e]\!]s' \in V$ which completes the proof. ∎

$$skip : (pts, lsh) \to (pts, lsh) \hookrightarrow skip$$

$$\frac{x \notin lsh \quad x := e : (pts, lsh) \to (pts', lsh)}{x := e : (pts, lsh) \to (pts', lsh) \hookrightarrow skip} (ass_1^d) \quad \frac{x \in lsh' \quad x := e : (pts, lsh) \to (pts', lsh')}{x := e : (pts, lsh) \to (pts', lsh') \hookrightarrow x := e} (ass_2^d)$$

$$\frac{I = \emptyset}{x := cons(e_1, \ldots, e_n) : (pts, lsh) \to (pts', lsh') \hookrightarrow x := cons(0_1, \ldots, 0_n)} (con_1^d) \quad \frac{I \neq \emptyset}{x := cons(e_1, \ldots, e_n) : (pts, lsh) \to (pts', lsh') \hookrightarrow x := cons(e_1, \ldots, e_n)} (con_2^d)$$

$$\frac{x \notin lsh \quad x := [e] : (pts, lsh) \to (pts', lsh)}{x := [e] : (pts, lsh) \to (pts', lsh) \hookrightarrow skip} (lok_1^d) \quad \frac{x \in lsh' \quad x := [e] : (pts, lsh) \to (pts', lsh')}{x := [e] : (pts, lsh) \to (pts', lsh') \hookrightarrow x := [e]} (lok_2^d)$$

$$dispose(e) : (pts, lsh) \to (pts, lsh') \hookrightarrow dispose(e) \quad (dis^d)$$

$$\frac{S_1 : (pts, lsh) \to (pts'', lsh'') \hookrightarrow S'_1 \quad S_2 : (pts'', lsh'') \to (pts', lsh') \hookrightarrow S'_2}{S_1 ; S_2 : (pts, lsh) \to (pts', lsh') \hookrightarrow S'_1 ; S'_2}$$

$$\frac{V \cap lsh' = \emptyset}{[e_1] := e_2 : (pts, lsh) \to (pts', lsh') \hookrightarrow skip} (mut_1^d) \quad \frac{V \cap lsh' \neq \emptyset}{[e_1] := e_2 : (pts, lsh) \to (pts', lsh') \hookrightarrow [e_1] := e_2} (mut_2^d)$$

$$\frac{b : lsh \to lsh'' \quad S_t : (pts, lsh'') \to (pts', lsh') \hookrightarrow S'_t \quad S_f : (pts, lsh'') \to (pts', lsh') \hookrightarrow S'_f}{if\ b\ then\ S_t\ else\ S_f : (pts, lsh) \to (pts', lsh') \hookrightarrow if\ b\ then\ S'_t\ else\ S'_f} \quad \frac{b : lsh \to lsh' \quad S_t : (pts, lsh') \to (pts, lsh) \hookrightarrow S'_t}{while\ b\ do\ S_t : (pts, lsh) \to (pts, lsh') \hookrightarrow while\ b\ do\ S'_t}$$

$$\frac{(pts'_1, lsh'_1) \leq (pts_1, lsh_1) \quad S : (pts_1, lsh_1) \to (pts_2, lsh_2) \hookrightarrow S' \quad (pts_2, lsh_2) \leq (pts'_2, lsh'_2)}{S : (pts'_1, lsh'_1) \to (pts'_2, lsh'_2) \hookrightarrow S'}$$

Fig.6. Inference rules of a type system for dead-code elimination.

**Theorem 2.** *1. $(pts, lsh) \leq (pts', lsh') \Rightarrow (\forall (s, h). (s,h) \models_{lsh} pts \Rightarrow (s, h) \models_{lsh'} pts')$.*
*2. Suppose that $S: (pts, lsh) \to (pts', lsh')$ and $S : (s, h) \to (s', h')$. Then $(s, h) \models_{lsh} pts$ implies $(s', h') \models_{lsh'} pts'$.*
*3. Suppose $S: (s,h) \to (s',h')$ and $S : (pts,lsh) \to (pts', lsh')$. Then $(s,h) \models lsh$ implies $(s',h') \models lsh'$. This guarantees that if the set of variables and memory locations live at the state $(s',h')$ is included in lsh', then the set of variables and memory locations live at the state $(s, h)$ is included in lsh.*
*Proof.* 1. Suppose $(s, h) \models_{lsh} pts$. This implies $(s,h) \models_{lsh'} pts$ because $lsh' \subseteq lsh$. The last fact implies $(s, h) \models_{lsh'} pts'$ because $pts \leq pts'$.
2. The proof is by induction on the structure of type derivation as follows:
(a) The type derivation has the form $(dis^l)$. In this case, $lsh = lsh' \cup FV(e)$ and $(s', h') = (s, h](dom(h) \setminus \{[\![e]\!]s\}))$. Therefore $(s,h) \models_{lsh} pts$ implies $(s',h) \models_{lsh} pts'$ which implies $(s', h') \models_{lsh'} pts'$ because $h' \leq h$ and $lsh' \subseteq lsh$.
(b) The type derivation has the form $(ass^l)$. In this case, $pts' = [pts \mid x : V']$ and $(s', h') = ([s \mid x : [\![e]\!]s], h)$. Therefore $(s, h) \models_{lsh} pts$ implies $(s', h) \models_{lsh} pts'$ because $x \notin lsh$. Clearly $(s', h) \models_{lsh} pts'$ implies $(s', h') \models_{lsh'} pts'$.

(c) The type derivation has the form $(ass_2^l)$. In this case, $pts' = [pts \mid x:V']$, $(s', h') = ([s \mid x : [\![e]\!]s], h)$, and $lsh = (lsh' \setminus \{x\}) \cup FV(e)$. Therefore by Lemma 4 and similarly to Theorem 1 (2.a), we can conclude $(s', h') \models_{lsh'} pts'$.

(d) The type derivation has the form $(con_1^l)$. In this case,
- $pts' = \cup_{1 \leq i \leq v} [pts \mid x: \{a_{n,1}^i\} \mid a_{n,1}^i : V'_1 \mid \ldots \mid a_{n,n}^i : V'_n]$,
- $(s', h') = ([s \mid x: a_{n,1}^u], [h \mid a_{n,1}^u : [\![e_1]\!]s \mid \ldots \mid a_{n,n}^u : [\![e_n]\!]s])$,
- $v = min\{t \mid \{a_{n,1}^t, \ldots, a_{n,n}^t\} \cap dom(pts) = \emptyset\}$, and
- $u = min\{t \mid \{a_{n,1}^t, \ldots, a_{n,n}^t\} \cap dom(h) = \emptyset\}$.

By Lemma 2, $1 \leq u \leq v$. Because $I = \emptyset$, $(s, h) \models_{lsh'} pts$ implies $(s', h') \models_{lsh'} pts'$.

(e) The type derivation has the form $(con_2^l)$. In this case, $(s', h')$, $pts'$, $u$, and $v$ have the same definitions as in the previous case (d). Moreover, $lsh = lsh' \setminus \{x\}$. Because $I = \{x\}$ and $1 \leq u \leq v$, $(s, h) \models_{lsh' \setminus \{x\}} pts$ implies $(s', h') \models_{lsh'} pts'$.

(f) The type derivation has the form $(con_3^l)$. In this case, $(s', h')$, $pts'$, $u$, and $v$ have the same definitions as in the case (d). Moreover $lsh = \cup_{a_{j,n,i} \in I} lsh_i$. Therefore by Lemma 4 and similarly to Theorem 1 (2.c), we can conclude $(s', h') \models_{lsh'} pts'$.

(g) The type derivation has the form $(lok_1^l)$. In this case, $pts' = [pts \mid x : \cup_{a \in V'} pts(a)]$, $(s', h') = ([s \mid x : h([\![e]\!]s)], h))$, and $lsh = lsh'$. Therefore $(s, h) \models_{lsh} pts$ implies $(s', h') \models_{lsh} pts'$ because $x \notin lsh$. Clearly $(s', h') \models_{lsh} pts'$ implies $(s', h') \models_{lsh'} pts'$.

(h) The type derivation has the form $(lok_2^l)$. In this case, $pts' = [pts \mid x : \cup_{a \in V'} pts(a)]$, $(s', h') = ([s \mid x : h([\![e]\!]s)], h)$, and $lsh = (lsh' \setminus \{x\}) \cup FV(e) \cup V'$. Therefore by Lemma 4 and similarly to Theorem 1 (2.b), we can conclude $(s', h') \models_{lsh'} pts'$.

(i) The type derivation has the form $(mut_1^l)$. In this case, $pts' = \cup_{a \in V'_1} [pts \mid a: V'_2]$, $(s', h') = (s, [h \mid [\![e_1]\!]s: [\![e_2]\!]s])$, and $lsh = lsh' \cup FV(e_1)$. Clearly, $(s, h) \models_{lsh} pts$ implies $(s, h) \models_{lsh'} pts$. Because $V \cap lsh' = \emptyset$, $(s, h) \models_{lsh'} pts$ implies $(s', h') \models_{lsh'} pts'$.

(j) The type derivation has the form $(mut_2^l)$. In this case, $pts' = \cup_{a \in V'_1} [pts \mid a: V'_2]$, $(s', h') = (s, [h \mid [\![e_1]\!]s: [\![e_2]\!]s])$, and $lsh = lsh' \cup FV(e_1) \cup FV(e_2)$. Therefore by Lemma 4 and similarly to Theorem 1 (2.d), we can conclude $(s', h') \models_{lsh'} pts'$.

The remaining cases are straightforward to check.

3. The proof again is by induction on the structure of type derivation and it is straightforward. ∎

**Corollary 1.** *Suppose* $S: (s, h) \rightarrow (s', h')$ *and* $S : (pts, lsh) \rightarrow (pts', lsh')$. *Then* $(s, h) \models (pts, lsh)$ *implies* $(s', h') \models (pts', lsh')$.

**Proof.** The proof follows from Theorem 2.

**Theorem 3.** *Suppose that* $S : (pts, lsh) \rightarrow (pts', lsh')$, $S: (s, h) \rightarrow (s', h')$, *and* $(s, h) \sim_{(pts,lsh)} (s_*, h_*)$. *Then there exists a state* $(s'_*, h'_*)$ *such that* $S: (s_*, h_*) \rightarrow (s'_*, h'_*)$ *and* $(s', h') \sim_{(pts',lsh')} (s'_*, h'_*)$.

**Proof.** The proof is by induction on structure of type derivation as follows:

1. The type derivation has the form $(dis^l)$. In this case, $lsh = lsh' \cup FV(e)$ and $(s', h') = (s, h](dom(h) \setminus \{[\![e]\!]s\}))$. Take $(s'_*, h'_*) = (s_*, h_*](dom(h_*) \setminus \{[\![e]\!]s_*\}))$.

2. The type derivation has one of the forms $(ass_1^l)$ and $(ass_2^l)$. In this case, $pts' = [pts \mid x : V']$ and $(s', h') = ([s \mid x : [\![e]\!]s], h)$. Take $(s'_*, h'_*) = ([s_* \mid x : [\![e]\!]s_*], h_*)$.

3. The type derivation has one of the forms $(con_1^l)$, $(con_2^l)$, and $(con_3^l)$. In this case,
- $pts' = \cup_{1 \leq i \leq v}[pts \mid x : \{a_{n,1}^i\} \mid a_{n,1}^i : V'_1 \mid \ldots \mid a_{n,n}^i : V'_n]$,
- $(s', h') = ([s \mid x : a_{n,1}^u], [h \mid a_{n,1}^u : [\![e_1]\!]s \mid \ldots \mid a_{n,n}^u : [\![e_n]\!]s])$,
- $v = min\{t \mid \{a_{n,1}^t, \ldots, a_{n,n}^t\} \cap dom(pts) = \emptyset\}$, and
- $u = min\{t \mid \{a_{n,1}^t, \ldots, a_{n,n}^t\} \cap dom(h) = \emptyset\}$.

By Lemma 2, $1 \leq u \leq v$. Take $(s'_*, h'_*) = ([s_* \mid x: a_{n,1}^u], [h_* \mid a_{n,1}^u : [\![e_1]\!]s_* \mid \ldots \mid a_{n,n}^u : [\![e_n]\!]s_*])$.

4. The type derivation has the form $(lok_2^l)$ or $(lok_1^l)$. In this case, $pts' = [pts \mid x : \cup_{a \in V'} pts(a)]$ and $(s', h') = ([s \mid x : h([\![e]\!]s)], h))$. Take $(s'_*, h'_*) = ([s_* \mid x: h_*([\![e]\!]s_*)], h))$.

5. The type derivation has one of the forms $(mut_1^l)$ and $(mut_2^l)$. In this case, $pts' = \cup_{a \in V'_1} [pts \mid a: V'_2]$ and $(s', h') = (s, [h \mid [\![e_1]\!]s: [\![e_2]\!]s])$. Take $(s'_*, h'_*) = (s_*, [h_* \mid [\![e_1]\!]s_*: [\![e_2]\!]s_*])$

The remaining cases are straightforward. ∎

## 5. Dead-code elimination

A type system for dead-code elimination is presented in this section. In a program, statements that have no effect on the content of variables and memory locations live at the end of the program are considered to be dead code. It is the task of the type system presented here to optimize programs via eliminating dead code. If the dead code is faulty (causing the program to abort), then removing it may result in correcting the program.

A typical judgment of our type system takes the form $S: (pts, lsh) \rightarrow (pts', lsh') \hookrightarrow S'$. And it implies that $S'$ optimizes $S$ towards dead-code elimination (and may be program correction). The derivation of a judgment provides a justification for the optimization process. It is clear from the form of the judgment that the optimization process is built on the type information gathered by our type system for live stack-heap analysis.

---

Algorithm: *optimize*
- Input: a statement $S$ of the language $while_p$ and a set of variables $lsh'$ that we like to consider live (have interest in their values) at the end of executing $S$;
- Output: an optimized and may be corrected version $S'$ of $S$ such that the relation between $S$ and $S'$ is as stated in Theorem 4.
- Method:
1. Find $pts$ such that $S: \bot \rightarrow pts$ in our type system for pointer analysis.
2. Find $lsh$ such that $S: (\bot, lsh) \rightarrow (pts, lsh')$ in our type system for live stack-heap analysis.
3. Find $S'$ such that $S: (\bot, lsh) \rightarrow (pts, lsh') \hookrightarrow S'$ in our type system for dead code elimination.

Fig.7. the algorithm *optimize*

The optimization process can be summarized in the algorithm *optimize* out-lined in Figure 7. The first step of the algorithm *optimize* annotates the program points of $S$ with pointer information. This is done via a post type derivation of $S$ for the bottom points-to-type ⊥ as the pre type. The second step of the algorithm refines the information obtained in the first step via annotating the program points with information about live variables and memory locations. This is done via a pre type derivation of $S$ for the set $lsh'$, the set of variables whose values concerns us at the end of execution, as the post type. Finally the information calculated in the second step is used in the third step to find $S'$ using our type system for dead code elimination.

The inference rules of our type system for dead-code elimination are presented in Figure 6. In the following rules for allocation, the set $I$ has the same definition as in the previous type system and we suppose that $x:=cons(e_1, \ldots, e_n) : (pts, lsh) \rightarrow (pts', lsh')$. In rules for mutation, we suppose that $[e_1]:= e_2: (pts, lsh) \rightarrow (pts', lsh')$ and $e_1 : pts \rightarrow V$. In rules for dispose, we suppose that $dispose(e) : (pts, lsh) \rightarrow (pts', lsh')$ and $e : pts \rightarrow V$.

We note that the rule $(con^d_1)$ transforms the allocation statement to $x:=cons(0, \ldots, 0)$, with $n$ arguments, rather than to *skip*. This is so because the optimization to *skip* would led to the possibility that sequent allocations allocates different arrays in the original and optimized version of the program. And this in turn would complicate the definition of similarity between states (Definition 5) used in proving the equivalence of the original and optimized version of the program. This complication does not worth introducing the *skip* statement. However a simple extra forward traverse of the program can remove all such allocations ($cons(0, \ldots, 0)$) if necessary.

The following theorem guarantees that if the original and optimized programs are executed in similar states and the original one does not abort then:
1. the optimized program does not abort as well, and
2. the optimized program reaches a state similar to that reached by the original program.

**Theorem 4.** (*Soundness*)
Suppose that $S:(pts, lsh) \rightarrow (pts', lsh') \hookrightarrow S'$ and $(s, h) \sim_{(lsh,pts)} (s_*, h_*)$. Then
1. If $S: (s, h) \rightarrow (s', h')$, then there exists a state $(s'_*, h'_*)$ such that $S': (s_*, h_*) \rightarrow (s'_*, h'_*)$ and $(s', h') \sim_{(pts',lsh')} (s'_*, h'_*)$.
2. If $S': (s_*, h_*) \rightarrow (s'_*, h'_*)$ and $S$ does not abort at $(s, h)$, then there exists a state $(s', h')$ such that $S: (s, h) \rightarrow (s', h')$ and $(s', h') \sim_{(pts',lsh')} (s'_*, h'_*)$.

*Proof.* 1. The proof is by induction on structure of type derivation as follows:
(a) The type derivation has the form $(ass_1^d)$. In this case, $S' = skip$. Take $(s'_*, h'_*) = (s_*, h_*)$.
(b) The type derivation has the form $(ass_2^d)$. In this case, $S' = S$. This case follows from Theorem 3.
(c) The type derivation has the form $(con_1^d)$. In this case, $S' = cons(0_1, \ldots, 0_n)$. Take $(s'_*, h'_*) = ([s_* | x: a^u_{n,1}], [h_* | a^u_{n,1} : 0 | \ldots | a^u_{n,n} : 0])$.
(d) The type derivation has the form $(con_2^d)$. In this case, $S' = S$. This case follows from Theorem 3.
(e) The type derivation has the form $(lok_1^d)$. In this case, $S' = skip$. Take $(s'_*, h'_*) = (s_*, h_*)$.
(f) The type derivation has the form $(lok_2^d)$. In this case, $S' = S$. This case follows from Theorem 3.
(g) The type derivation has form $(mut_1^d)$. In this case, $S' = skip$. Take $(s'_*, h'_*) = (s_*, h_*)$.
(h) The type derivation has the form $(mut_2^d)$. In this case, $S' = S$. This case follows from Theorem 3.
The remaining cases are straightforward.
2. Similar to 1. ∎

## 6. Related work

There are two fields of related work; the first is type systems for data-flow analysis and the second is pointer analysis for sequential languages.

In [9] it is shown that a good deal of static analysis can be done in the type-systems fashion. More precisely, for every analysis in a certain class of data-flow analyses, it is proved in [9] that there exists a type system such that a program checks with a type if and only if the type is a super-type for the set resulting from running the analysis on the program. Later on [20], based on [9], established compositional type systems to carry program and proof optimization based on dead-code and common sub-expression elimination for the while language. These type systems are equipped with a transformation component that does the actual optimization. Our paper builds on and extends results presented in [20] to pointer languages.

The type system in [11] and the flow-logic work in [15] (used in [12, 13] to study security of the coordinated systems) are very similar to [9]. [2] presents constant folding and dead-code elimination via type systems and also introduces relational Hoare logic used to prove correctness of optimizations. Type systems for bidirectional data-flow analyses and their program optimizations are presented in [6]. Earlier, related work (with type systems that are structurally-complex) is [16, 17]. However none of these papers consider pointer programs.

Pointer analysis for *C*−like programs has been actively studied for decades [10, 23, 22, 7, 1, 4, 5, 8, 21, 3]. However none of these papers utilize results of their pointer analyses in data-flow analyses resulting in program optimization and/or correction associated with a justification for the transformation. The objective of pointer analysis has been to obtain a sound analysis only for the sake of program transformation and/or understanding. In [21], the bi-similarity is used to find pointer equivalences in a technique optimizing the performance of inclusion-based pointer analysis. But this does not go farther by showing how the task of optimizing programs is affected. In [3], a conditionally-sound pointer analysis is presented. The results of this analysis are utilized towards checking memory safety, but again the result of the verification is not associated with a justification. [10] studies the existence of an equivalent to shape analysis for purely functional programs, and the "shapes" it discovers. The argument in [10] is that by treating binding environments as dynamically allocated structures, by treating bindings as addresses, and by treating value environments as heaps, the "shape" of higher-order functions can be analyzed. The better your paper looks, the better the Journal looks. Thanks for your cooperation and contribution.

## Acknowledgments

This work was started during the author's sabbatical at Institute of Cybernetics, Estonia in the year 2009. The author is grateful to T. Uustalu for fruitful discussions. This work was partially supported by the EU FP6 IST project MOBIUS.

**Mohamed A. El-Zawawy** received: PhD in Computer Science from the University of Birmingham in 2007, M.Sc. in Computational Sciences in 2002 from Cairo University, and a BSc. in Computer Science in 1999 from Cairo University. Dr El-Zawawy is an assistant professor of Computer Science at Faculty of Science, Cairo University Since 2007. During the year 2009, Dr. El-Zawawy held the position of an extra-ordinary senior research at the Institute of Cybernetics, Tallinn University of Technology, Estonia. Dr. El-Zawawy worked as a teaching assistant at Cairo University from 1999 to 2003 and latter at Birmingham University from 2003 to 2007. Dr. El-Zawawy is interested in static analysis, shape analysis, type systems, and semantics of programming languages.